\documentstyle[a4,12pt]{article}
\newcommand{\beq}{\begin{equation}}
\newcommand{\eeq}{\end{equation}}
\newcommand{\beqa}{\begin{eqnarray}}
\newcommand{\eeqa}{\end{eqnarray}}
\newcommand{\no}{\nonumber}

\newcommand{\ol}{\overline}
\newcommand{\ra}{\rightarrow}

\newcommand{\ve}{\varepsilon}
\newcommand{\vp}{\varphi}

\newcommand{\dg}{\dagger}
\newcommand{\wt}{\widetilde}

\newcommand{\dfrac}{\displaystyle \frac}
\newcommand{\dis}{\displaystyle}

\newcommand{\ben}{\begin{enumerate}}
\newcommand{\een}{\end{enumerate}}
\newcommand{\bfl}{\begin{flushleft}}
\newcommand{\efl}{\end{flushleft}}
\newcommand{\ba}{\begin{array}}
\newcommand{\ea}{\end{array}}
\newcommand{\btab}{\begin{tabular}}
\newcommand{\etab}{\end{tabular}}
\newcommand{\bit}{\begin{itemize}}
\newcommand{\eit}{\end{itemize}}

\normalsize
\begin{document}
\parskip=4pt plus 1pt  
\textheight=8.7in
\bibliographystyle{unsrt}
\begin{titlepage}
\begin{flushright}
CERN-TH.6444/92\\
\end{flushright}
\vspace{1.5cm}
\begin{center}
{\Large \bf
THE CHIRAL ANOMALY \\[5pt]
IN NON-LEPTONIC WEAK INTERACTIONS $^*$} \\[60pt]
{\bf J. Bijnens$^1$, G. Ecker$^{1,2}$ and A. Pich$^1$} \\[20pt]
$^{1)}$  CERN, CH-1211 Geneva 23, Switzerland\\
$^{2)}$  Inst. Theor. Physik, Univ. Wien, Boltzmanng. 5,
Vienna, Austria \\[80pt]
{\bf ABSTRACT} \\[10pt]
\end{center}
\noindent
The interplay between the chiral anomaly and the non-leptonic weak
Hamiltonian is studied. The structure of the corresponding effective
Lagrangian of odd intrinsic parity is established. It is shown that the
factorizable contributions (leading in $1/N_C$) to that
Lagrangian can be calculated without free parameters. As a first
application, the decay $K^+ \ra \pi^+ \pi^0 \gamma$ is investigated.

\vfill
\begin{enumerate}
\item[*)] Work supported in part by CICYT (Spain), Grant No. AEN90-0040.
\end{enumerate}
\begin{flushleft}
CERN-TH.6444/92\\
April 1992
\end{flushleft}
\end{titlepage}

\paragraph{1.} The chiral anomaly \cite{ABJB} is a fundamental property
of chiral quantum field theories such as the standard model.
Although its origin as an intrinsically quantum mechanical violation of a
classical symmetry is well understood, many aspects of the anomaly remain
to be tested experimentally. For the strong, electromagnetic and
semileptonic weak interactions, the manifestations of the chiral anomaly
at low energies are completely determined by the Wess--Zumino--Witten
(WZW) functional \cite{WZW} in terms of pseudoscalar meson and
external gauge fields.

The non-leptonic weak interactions, which are the subject of this letter,
require a separate treatment. This can already be seen in the normal
parity sector. To lowest order in chiral perturbation theory (CHPT), the
strong, electromagnetic and semileptonic weak interactions of pseudoscalar
mesons are governed by the effective chiral Lagrangian of $O(p^2)$
(in the notation of Ref. \cite{GLNP})
\beq
{\cal L}_2 = \frac{F^2}{4} \langle D_\mu U D^\mu U^\dg +
             \chi U^\dg + \chi^\dg U \rangle, \label{eq:L2} \eeq
where
\beq
D_\mu U = \partial_\mu U - ir_\mu U + iU l_\mu, \qquad
\chi = 2 B_0(s + ip), \eeq
and $\langle A \rangle$ stands for the trace of the matrix $A$; $U$ is a
unitary $3 \times 3$ matrix
$$
U^\dg U = {\bf 1}, \qquad \det U = 1,
$$
which transforms as
\beq
U \ra g_R U g^\dg_L \eeq
under $SU(3)_L \times SU(3)_R$ and
incorporates the eight pseudoscalar Goldstone boson fields.
The external $3\times 3$ hermitian matrix fields $l_\mu, r_\mu, s, p$
contain
in particular the relevant gauge fields of the standard model
for electromagnetic and semileptonic weak interactions
\beqa
r_\mu  =  v_\mu + a_\mu & = & eQ A_\mu
\label{eq:gf} \\*
l_\mu  =  v_\mu - a_\mu & = & eQ A_\mu +
\dfrac{e}{\sqrt{2}\sin{\theta_W}} (W^+_\mu T_+ + h.c.) \no
\eeqa $$
Q = \dfrac{1}{3}~\rm{diag}(2,-1,-1)$$
$$
T_+ = \left( \ba{ccc}
0 & V_{ud} & V_{us} \\
0 & 0 & 0 \\
0 & 0 & 0 \ea \right)  $$
where the $V_{ij}$ are Kobayashi--Maskawa matrix elements.
The parameters $F$ and $B_0$ in the non-linear sigma model Lagrangian
(\ref{eq:L2}) are related to the pion decay constant
($F \simeq F_\pi = 93.2 MeV$)
and to the quark condensate, respectively \cite{GLNP}.

Although the Lagrangian (\ref{eq:L2})
allows one in particular to calculate
any mesonic amplitude with an external $W$ to $O(p^2)$, one does
not obtain the full non-leptonic mesonic amplitudes at low energies by
simply contracting the $W$ field.
Instead, one first has to integrate out the $W$
together with the heavy quarks in the fundamental theory to arrive at an
effective $\Delta S = 1$ Hamiltonian \cite{GW}
\beq
{\cal H}_{eff}^{\Delta S = 1} = \dfrac{G_F}{\sqrt{2}} V_{ud} V_{us}^*
\sum_i C_i(\mu^2) Q_i + {\rm h.c.}    \label{eq:Heff} \eeq
The Wilson coefficients $C_i(\mu^2)$ are functions of the heavy masses,
$\Lambda_{QCD}$ and the renormalization scale $\mu$. The $Q_i$ are
the standard four-quark operators which can be written as
products of colour singlet quark bilinears.

Restricting ourselves to the dominant octet part, the effective
Hamiltonian (\ref{eq:Heff})
has a unique realization at the mesonic level to lowest order in CHPT
first given by Cronin \cite{Cronin}
\beq
{\cal L}_2^{\Delta S = 1} = G_8 F^4 \langle \lambda D_\mu U^\dg
D^\mu U \rangle + {\rm h.c.} \label{eq:L2weak} \eeq
$$ \lambda = \dfrac{1}{2}(\lambda_6 - i \lambda_7). $$
The only coupling constant at $O(p^2)$ can be determined from $K \ra
2 \pi$ decays to be
\beq
|G_8| \simeq 9\times 10^{-6} \mbox{ GeV}^{-2} \simeq
5 \times \dfrac{G_F}{\sqrt{2}} |V_{ud} V_{us}|. \eeq
$G_8$ exhibits the non-leptonic enhancement factor
($\Delta I = 1/2$ rule), but is subject to large higher-order
corrections \cite{KMW2}.

\paragraph{2.}
The chiral anomaly enters at $O(p^4)$. Its contribution to
strong, electromagnetic and semileptonic weak amplitudes is contained
in the Wess--Zumino--Witten functional \cite{WZW},
which has the following
explicit form in a scheme where the vector currents are conserved:
\beqa
S[U,l,r]_{WZW} &=&-\dfrac{i N_C}{240 \pi^2}
\int d\sigma^{ijklm} \langle \Sigma^L_i
\Sigma^L_j \Sigma^L_k \Sigma^L_l \Sigma^L_m \rangle \label{eq:WZW} \\*
 & & - \dfrac{i N_C}{48 \pi^2} \int d^4 x
\varepsilon_{\mu \nu \alpha \beta}\left( W (U,l,r)^{\mu \nu
\alpha \beta} - W ({\bf 1},l,r)^{\mu \nu \alpha \beta} \right)
\no \\
W (U,l,r)_{\mu \nu \alpha \beta} & = &
\bigl \langle U l_{\mu} l_{\nu} l_{\alpha}U^{\dg} r_{\beta}
+ \frac{1}{4} U l_{\mu} U^{\dg} r_{\nu} U l_\alpha U^{\dg} r_{\beta}
+ i U \partial_{\mu} l_{\nu} l_{\alpha} U^{\dg} r_{\beta}
\no  \\
& & +~ i \partial_{\mu} r_{\nu} U l_{\alpha} U^{\dg} r_{\beta}
- i \Sigma^L_{\mu} l_{\nu} U^{\dg} r_{\alpha} U l_{\beta}
+ \Sigma^L_{\mu} U^{\dg} \partial_{\nu} r_{\alpha} U l_\beta
\no \\
& & -~ \Sigma^L_{\mu} \Sigma^L_{\nu} U^{\dg} r_{\alpha} U l_{\beta}
+ \Sigma^L_{\mu} l_{\nu} \partial_{\alpha} l_{\beta}
+ \Sigma^L_{\mu} \partial_{\nu} l_{\alpha} l_{\beta}  \\
& & -~ i \Sigma^L_{\mu} l_{\nu} l_{\alpha} l_{\beta}
+ \frac{1}{2} \Sigma^L_{\mu} l_{\nu} \Sigma^L_{\alpha} l_{\beta}
- i \Sigma^L_{\mu} \Sigma^L_{\nu} \Sigma^L_{\alpha} l_{\beta}
\bigr \rangle \no \\
& & -~ \left( L \leftrightarrow R \right) \no \eeqa
$$
\Sigma^L_\mu = U^{\dg} \partial_\mu U \qquad
\Sigma^R_\mu = U \partial_\mu U^{\dg} $$
$$ N_C = 3 \qquad \varepsilon_{0123} = 1  $$
where $\left( L \leftrightarrow R \right)$ stands for the interchange
$$
U \leftrightarrow U^\dg, \qquad l_\mu \leftrightarrow r_\mu,
\qquad \Sigma^L_\mu \leftrightarrow \Sigma^R_\mu . $$

The anomaly also contributes to non-leptonic weak amplitudes starting at
$O(p^4)$. The most obvious contribution is due to tree diagrams
involving one WZW vertex and one vertex from
the non-leptonic weak Lagrangian (\ref{eq:L2weak}). Since
${\cal L}_2^{\Delta S = 1}$ contains bilinear terms in the meson fields,
there is a local part in the corresponding functional which can be given
in explicit form by diagonalizing the kinetic part of the Lagrangians
(\ref{eq:L2}) and (\ref{eq:L2weak}) simultaneously \cite{EPR3}.
This local Lagrangian embodies the so-called pole
contributions to anomalous non-leptonic weak amplitudes and is given
by \cite{ENP1}
\beq
{\cal L}^{\Delta S =1}_{an} = - \frac{ieG_8}{8\pi^2 F} \wt F^{\mu\nu}
\partial_\mu \pi^0 K^+ \stackrel{\leftrightarrow}{D_\nu} \pi^-
+ \frac{\alpha G_8}{6\pi f} \wt F^{\mu\nu} F_{\mu\nu}
\left( K^+ \pi^- \pi^0 - \frac{1}{\sqrt{2}} \, K^0 \pi^+ \pi^-\right)
+ {\rm h.c.}
\label{eq:Lan} \eeq
$F_{\mu\nu} = \partial_\mu A_\nu - \partial_\nu A_\mu$ is
the electromagnetic field strength tensor,
$\wt F_{\mu\nu} = \ve_{\mu\nu\rho\sigma} F^{\rho\sigma}$
its dual, and $D_\nu \vp^\pm$ denotes the covariant derivative
$(\partial_\nu \mp ieA_\nu)\vp^\pm$. In the limit of CP conservation,
the anomalous Lagrangian (\ref{eq:Lan}) contributes only
to the decays (with real or virtual photons)
\beq
K^+ \ra \pi^+ \pi^0 \gamma,\; \pi^+ \pi^0 \gamma \gamma \quad
{\rm and} \quad K_L \ra \pi^+ \pi^- \gamma \gamma. \label{eq:andec}
\eeq

There is, however, an additional source of non-leptonic anomalous
amplitudes of $O(p^4)$, which was not taken into account in
Ref. \cite{ENP1}. Diagrammatically, those contributions can be pictured
as arising from contraction of the $W$ field between Green functions due
to the anomaly on the one side and the Lagrangian ${\cal L}_2$
of Eq. (\ref{eq:L2}) on the other side.
However, as in the normal parity sector
discussed before, such a procedure would not give the correct amplitudes
at the hadronic scale. Rather, we must use again the operator product
expansion first and realize the corresponding operators at the bosonic
level in the presence of the anomaly.

\paragraph{3.}
A possible framework to implement the bosonization of four-quark
operators was formulated in Ref. \cite{PR}. Let us first recall the
standard bosonization of (left-handed) quark currents in CHPT. The Green
functions of quark currents can be expressed as functional integrals in
the fundamental theory
\beqa
\langle 0|T\{\ol{q_{jL}}\gamma^\mu q_{iL}\dots \}|0\rangle & = &
N^{-1} \int [DqD\ol{q}DG] \ol{q_{jL}}\gamma^\mu q_{iL}\dots
e^{\dis i \int d^4x {\cal L}} \no \\*
& = & (- i \dfrac{\delta}{\delta l_{\mu,ji}}) \dots
\rm e^{\dis i Z[l,r,s,p]}
\eeqa
where \cite{GLNP}
\beq
{\cal L} = {\cal L}^0_{QCD} + \ol{q}\gamma^\mu \{v_\mu + \gamma_5
a_\mu\}q - \ol{q}\{s - i \gamma_5 p\}q
\eeq
is the QCD Lagrangian with massless light quarks in the presence of
external fields $l,r,s,p$. The basic tenet of CHPT is that the
generating functional $Z[l,r,s,p]$ \cite{GLNP}
can be calculated in the effective theory in terms of the
chiral effective action $S[U,l,r,s,p]$ as
\beq
\rm e^{\dis i Z[l,r,s,p]} = N^{-1} \int [DU]\rm e^{\dis i S[U,l,r,s,p]}
\ .\eeq
The bosonized form of the quark currents in the low-energy theory
is then given by
\beq
\ol{q_{jL}}\gamma^\mu q_{iL} \leftrightarrow \dfrac{\delta S[U,l,r,s,p]}
{\delta l_{\mu,ji}}. \label{eq:current} \eeq

The insertion of a four-quark operator $Q_i$ can be treated in a similar
fashion \cite{PR}. In the effective chiral theory, a four-quark operator
\footnote{We restrict the analysis to products of left-chiral currents.}
corresponds to the insertion \cite{PR}
\beq
\langle 0|T\{\ol{q_{lL}}\gamma^\mu q_{kL}\ol{q_{jL}}\gamma_\mu q_{iL}
\dots \}|0\rangle = N^{-1}\int [DUDG] \left\{\dfrac{\delta\Gamma}{\delta
l_{\mu,lk}}\dfrac{\delta\Gamma}{\delta l^\mu_{ji}} - i \dfrac{\delta^2
\Gamma}{\delta l_{\mu,lk}\delta l^\mu_{ji}}\right\}\dots
\rm e^{\dis i \Gamma} \label{eq:4q} \eeq
where $\Gamma[U,l,r,s,p;G]$ is the effective action before the gluons
are integrated out. There is a difference between the bosonization of
a current
(\ref{eq:current}) and a four-quark operator (\ref{eq:4q}) due to
the anomalous dimension of the four-quark operator. The gluonic integral
includes soft gluons dressing the four-quark operator [recall that only
hard gluons were integrated out to arrive at the Hamiltonian
(\ref{eq:Heff})] to produce the necessary $\mu$-dependence of the
operator in its bosonized form. Of course, neither $S[U,l,r,s,p]$ nor
$\Gamma[U,l,r,s,p;G]$ can be calculated directly from QCD at this time.
However, very encouraging progress in this direction has recently been
made in the context of a model incorporating a specific mechanism for
spontaneous chiral symmetry breaking \cite{ERT,PR,Bij}.

The two contributions in Eq. (\ref{eq:4q})
are denoted \cite{PR} as factorizable (leading
in $1/N_C$) and non-factorizable (non-leading in $1/N_C$), respectively.
Let us now consider the odd-intrinsic parity parts (containing the
$\ve$ tensor)
\beq
\Gamma^-[U,l,r,s,p;G] \qquad \mbox{and} \qquad S^-[U,l,r,s,p].
\eeq
There is obviously no term of $O(p^2)$ in $\Gamma^-$.
Nor does chiral symmetry allow a chiral invariant $O(p^4)$ term.
Instead, $\Gamma^-$ is given to $O(p^4)$ by the chiral anomaly in terms
of the WZW functional $S[U,l,r]_{WZW}$ and by a term accounting for the
gluonic component of the chiral $U(1)$ anomaly. Except for this latter
part, which will not play any special r\^ole in the following,  the
remainder $\Gamma^-_{rem}[U,l,r,s,p;G]$ defined via
\beq
\Gamma^-[U,l,r,s,p;G] = S[U,l,r]_{WZW} + \Gamma^-_{rem}[U,l,r,s,p;G]
\label{eq:Gamma} \eeq
is chiral-invariant. The anomaly can be viewed as arising
from the chiral non-invariance of the fermionic measure in
the path integral \cite{Fuji}. The non-renormalization theorem \cite{AB}
of the chiral anomaly then tells us that there is a similar decomposition
for the effective action $S^-$ with odd intrinsic parity,
\beq
\label{eq:SMINUS}
S^-[U,l,r,s,p] = S[U,l,r]_{WZW} + S^-_{inv}[U,l,r,s,p] .
\eeq
The $O(p^4)$ part of $S^-$ is unaffected by the gluonic path integral
and is again given by $S[U,l,r]_{WZW}$. The remainder
$S^-_{inv}[U,l,r,s,p]$ starts at $O(p^6)$ and is $SU(3)_L \times
SU(3)_R$ invariant \cite{DWBBCI}.

Looking at Eq. (\ref{eq:4q}), we find that the WZW functional in
Eq. (\ref{eq:Gamma})
only contributes to the factorizable part because there is no possible
contribution of $O(p^2)$ with an $\ve$ tensor. The non-factorizable
contribution is determined entirely by $\Gamma_{rem}^-$ and it can only
be calculated with a special model for spontaneous chiral symmetry
breaking (cf. Refs. \cite{ERT,PR,Bij}).

To lowest order, $O(p^4)$, the factorizable contribution can be given
exactly because the WZW functional can be pulled out of the gluonic
path integral. The bosonized form of the four-quark operator in the
anomalous parity sector is [factorizable contribution of $O(p^4)$] :
\beq
\ol{q_{lL}}\gamma^\mu q_{kL}\ol{q_{jL}}\gamma_\mu q_{iL}
\leftrightarrow \dfrac{\delta S_{WZW}}{\delta l^\mu_{lk}}\dfrac{\delta
S_2}{\delta l_{\mu,ji}} + \left(lk \leftrightarrow ji\right)
\label{eq:fact} \eeq
where
\beqa
\dfrac{\delta S_2}{\delta l_{\mu,ji}} & = & - \dfrac{F^2}{2}
\left(L^\mu\right)_{ij} \\
L^\mu & = & i U^\dg D^\mu U       \label{eq:curr2} \no
\eeqa
is the left-chiral current of lowest order $p$ corresponding to the
chiral Lagrangian (\ref{eq:L2}). The anomalous current [of $O(p^3)$]
has the following form
\beqa
\dfrac{\delta S_{WZW}}{\delta l_{\mu,ji}} & = & \dfrac{1}{16 \pi^2}
\ve^{\mu \nu \alpha \beta} J^{an}_{\nu \alpha \beta,ij} \no \\*
J^{an}_{\nu \alpha \beta} & = & i L_\nu L_\alpha L_\beta
+ \left\{F^L_{\nu \alpha} + \frac{1}{2}U^\dg
F^R_{\nu \alpha} U, L_\beta \right\} \label{eq:ancurr} \\*
& & + \mbox{ a chirally non-covariant polynomial}  \no \\*
& & \mbox{ in the external fields } l,r , \no
\eeqa
where $F^L, F^R$ are the non-Abelian field strengths associated with
the fields $l, r$ \cite{GLNP}.
The anomalous current (\ref{eq:ancurr}) has a well-known structure
\cite{BZ,Leu}:
it consists of a chirally covariant piece written explicitly
in (\ref{eq:ancurr}) and a local polynomial in the external gauge fields
$l,r$, which is not chirally covariant.
Only the covariant anomalous current has direct physical significance.
Changing the local polynomial in $l, r$ amounts to different
regularization schemes which cannot modify the physical content of
the anomaly. Moreover,
in the regularization scheme chosen in Eq. (\ref{eq:WZW}), the
local polynomial disappears for external vector gauge fields which is
exactly the case relevant for us: all external gauge fields are photons
(radiative non-leptonic kaon decays).

\paragraph{4.}
We can now construct the bosonization of the dominant octet operator in
(\ref{eq:Heff})
\beq
Q_- = Q_2 - Q_1 \eeq
$$
Q_1 = \ol{s}\gamma^\mu(1-\gamma_5)d\ol{u}\gamma_\mu(1-\gamma_5)u $$
$$
Q_2 = \ol{s}\gamma^\mu(1-\gamma_5)u\ol{u}\gamma_\mu(1-\gamma_5)d $$
in the factorizable approximation for the odd-parity part of $O(p^4)$.
The final result is
\beq
Q_-(\mbox{fact}) \leftrightarrow -\dfrac{F^2}{8 \pi^2}
\ve^{\mu\nu\alpha\beta}\left[\langle \lambda
\{L_\mu, J^{an}_{\nu \alpha \beta}\}\rangle - \langle \lambda
L_\mu\rangle \langle J^{an}_{\nu \alpha \beta}\rangle \right]
\label{eq:Q-1} \eeq
in terms of the covariant anomalous current (\ref{eq:ancurr}). This
result may be written in a more explicit form in terms of four chiral
operators of $O(p^4)$ of odd intrinsic parity
as \footnote{We neglect a term proportional to $\langle [\lambda,
\tilde{F}^{\mu\nu}_L] L_\mu L_\nu \rangle$,
which cannot contribute for the case of
external photons.}
\beqa
Q_-(\mbox{fact}) & \leftrightarrow & \dfrac{F^2}{16 \pi^2}
\left(2 i \ve^{\mu \nu \alpha \beta}\langle \lambda L_\mu \rangle
\langle L_\nu L_\alpha L_\beta \rangle \right. \no \\*
& & +~ \langle \lambda [U^\dg \tilde{F}^{\mu\nu}_R U, L_\mu L_\nu]
\rangle \no \\*
& & +~ 3 \langle \lambda L_\mu \rangle \langle (\tilde{F}^{\mu\nu}_L
+ U^\dg \tilde{F}^{\mu\nu}_R U) L_\nu \rangle \no \\*
& & + \left. \langle \lambda L_\mu \rangle \langle (\tilde{F}^{\mu\nu}_L
- U^\dg \tilde{F}^{\mu\nu}_R U) L_\nu \rangle \right) .
\label{eq:Q-2} \eeqa
The last three terms in Eq. (\ref{eq:Q-2}) were already obtained by
Cheng \cite{Cheng}, although not in the explicitly covariant form given
here. They were also discussed in Ref. \cite{ENP1} since they contribute
to the decays $K \ra \pi \pi
\gamma (\gamma)$ considered there. Contrary to the conjecture made
in Ref. \cite{ENP1}, the coefficients of these terms are by no means
small, since they are actually generated by the chiral anomaly although
the operators are completely chirally covariant. The first term in
Eq. (\ref{eq:Q-2}) only contributes to $K$ decays with at least three
pions in the final state. In fact, in the limit of CP conservation
it only contributes to the decay
$K_L \ra \pi^+ \pi^- \pi^0 \gamma$ \cite{ENP2}.

We have therefore established that the factorizable contribution to
$Q_-$ in the anomalous parity sector can be completely determined to
$O(p^4)$ as given by Eq. (\ref{eq:Q-2}). Moreover, of all possible
octet operators of $O(p^4)$ proportional to the $\ve$ tensor \cite{KMW1},
only the four operators appearing in (\ref{eq:Q-2}) can contribute
to processes where all external gauge fields are photons. Since the
WZW functional cannot contribute to the non-factorizable part,
the latter automatically has the right chiral transformation
property of an octet operator. Therefore, even if we cannot
calculate the non-factorizable contributions in a model-independent
way, we know they must be of the form (\ref{eq:Q-2}), albeit
with different coefficients. In fact, those coefficients will be
$\mu$-dependent to cancel the $\mu$-dependence of the Wilson
coefficients. This cancellation was discussed in Ref. \cite{PR}
for the normal-parity octet Lagrangian of $O(p^2)$ in (\ref{eq:L2weak}).

Since all octet operators in ${\cal H}^{\Delta S=1}_{eff}$ produce the
same structure (\ref{eq:Q-2}), we can write down the final
representation of the $\Delta S=1$ effective Lagrangian in the
anomalous parity sector to $O(p^4)$
\beqa
{\cal L}^{\Delta S=1}_{eff} & = & \dfrac{G_8 F^2}{16 \pi^2}
\left(2 a_1 i \ve^{\mu \nu \alpha \beta}\langle \lambda L_\mu \rangle
\langle L_\nu L_\alpha L_\beta \rangle \right. \no \\*
& & +~ a_2 \langle \lambda [U^\dg \tilde{F}^{\mu\nu}_R U, L_\mu L_\nu]
\rangle \no \\*
& & +~ 3 a_3 \langle \lambda L_\mu \rangle \langle (\tilde{F}^{\mu\nu}_L
+ U^\dg \tilde{F}^{\mu\nu}_R U) L_\nu \rangle \no \\*
& & + \left. a_4 \langle \lambda L_\mu \rangle \langle (\tilde{F}^
{\mu\nu}_L - U^\dg \tilde{F}^{\mu\nu}_R U) L_\nu \rangle \right) + h.c.
\label{eq:Hfinal} \eeqa
{}From the dominance of the octet operator $Q_-$ we expect the dimensionless
coefficients $a_i$ to be positive and of order one. In fact, they can be
expected to be somewhat smaller than one because there is no
factorizable contribution to (\ref{eq:Hfinal}) from the dominant
penguin operator $Q_6$ unlike the situation in the normal parity sector
at $O(p^2)$ \cite{PR}.

\paragraph{5.}
The phenomenological implications of the chiral anomaly for non-leptonic
$K$ decays will be treated in more detail elsewhere \cite{ENP2}. Here,
we confine ourselves to a brief application of our findings to the decay
$K^+ \ra \pi^+ \pi^0 \gamma$. The available experimental evidence is
consistent with a dominant magnetic part for the direct emission
amplitude (non-bremsstrahlung)\cite{K+}. In Ref. \cite{ENP1}, the
magnetic amplitude due to the Lagrangian (\ref{eq:Lan}) was found to be
\beq
M = - \dfrac{e G_8 M^3_K}{2 \pi^2 F} ,
\label{eq:Man1} \eeq
whereas (\ref{eq:Hfinal}) contributes
\beq
M = \dfrac{3 e G_8 M^3_K}{2 \pi^2 F}(\dfrac{a_2}{2} - a_3) .
\label{eq:Man2} \eeq
Moreover, it was shown in \cite{ENP1} that the $V$-exchange corrections
of $O(p^6)$ to $M$ are much smaller than the anomalous amplitude
(\ref{eq:Man1}) of $O(p^4)$.
For $a_i \simeq 1$, there is positive interference between the two
amplitudes \cite{Cheng}. Assuming that the direct emission rate is
purely magnetic, the experiments of Ref. \cite{K+} have found a
corresponding branching ratio (the average is from Ref. \cite{PDG})
\beq
BR(K^+ \ra \pi^+ \pi^0 \gamma)_{DE} = (1.8 \pm 0.4)\times 10^{-5}
\label{eq:BRexp} \eeq
for a certain cut in the kinetic energy of the charged pion. With the
total magnetic amplitude given by the sum of (\ref{eq:Man1}) and
(\ref{eq:Man2}), we find
\beq
BR(K^+ \ra \pi^+ \pi^0 \gamma)_M = 2.2\times 10^{-5}\left(\dfrac
{2 + 6a_3 -3a_2}{5}\right)^2(G_8/9\times 10^{-6}\mbox{ GeV}^{-2})^2
\label{eq:BRtheor} \eeq
for the same cuts. The good agreement between (\ref{eq:BRexp}) and
(\ref{eq:BRtheor}) gives support to the theoretical expectation
\beq
a_i \stackrel{<}{\sim} 1 .
\label{eq:ai} \eeq
Further experimental work isolating the competing electric direct
emission amplitude \cite{ENP1} is necessary to make this agreement
more conclusive.

\paragraph{Conclusions}
\ben
\item[i.] The non-Abelian chiral anomaly contributes to non-leptonic
weak processes in two different ways at $O(p^4)$:
\bit
\item CHPT amplitudes involving one WZW vertex and one vertex from
${\cal L}_2^{\Delta S=1}$ (\ref{eq:L2weak}). The so-called pole
contributions are given in closed form by the Lagrangian (\ref{eq:Lan}).
\item Direct anomalous amplitudes due to chirally covariant octet
operators of $O(p^4)$ in (\ref{eq:Q-2}) and (\ref{eq:Hfinal}). \eit
\item[ii.] In the non-leptonic sector, the anomaly contributes only
to radiative kaon decays.
\item[iii.] The factorizable contributions of odd intrinsic parity,
which are unambiguously calculable, already produce all possible terms
of $O(p^4)$ relevant for radiative processes where all external gauge
fields are photons. Due to the non-renormalization of the chiral
anomaly, there are no QCD corrections to the factorizable part
at $O(p^4)$.
\item[iv.] The non-factorizable terms (non-leading in $1/N_C$) compensate
the scale dependence of the Wilson coefficients. They contribute to the
coefficients of the factorizable terms but they cannot generate new ones.
\item[v.] Although the coefficients of the four possible chiral operators
are not calculable in a model-independent way, the overall scale is
expected to be of $O(G_8)$ or somewhat smaller. The octet enhancement of
the Wilson coefficients appears in this scale, but the
dominant penguin operator $Q_6$ does not contribute to the factorizable
part, because the WZW functional is independent of external
scalar/pseudoscalar fields.
\item[vi.] The phenomenological implications are encouraging for
the decay $K^+ \ra \pi^+ \pi^0 \gamma$ and will be considered in more
detail elsewhere \cite{ENP2}.
\een

\paragraph{Acknowledgements}
\paragraph{}
We thank J\"{u}rg Gasser and Heiri Leutwyler for clarifying discussions.

\end{document}